# Niobium superconducting nanowire single-photon detectors

Anthony J. Annunziata, Daniel F. Santavicca, Joel D. Chudow, Luigi Frunzio, Michael J. Rooks, Aviad Frydman, Daniel E. Prober

*Abstract*— We investigate the performance of superconducting nanowire photon detectors fabricated from ultra-thin Nb. A direct comparison is made between these detectors and similar nanowire detectors fabricated from NbN. We find that Nb detectors are significantly more susceptible than NbN to thermal instability (latching) at high bias. We show that the devices can be stabilized by reducing the input resistance of the readout. Nb detectors optimized in this way are shown to have approximately 2/3 the reset time of similar large-active-area NbN detectors of the same geometry, with approximately 6% detection efficiency for single photons at 470 nm.

## I. INTRODUCTION

In the past several years, there has been much interest in superconducting nanowires used as optical detectors [1] – [3]. These detectors, since their first implementation in NbN [1], have proven to be single photon sensitive with detection efficiency greater than 20% (as high as 75% using an optical resonator) with tens of picoseconds of jitter, low dark counts, and single photon count rates of ~100 MHz for visible and infrared photons up to 1550 nm wavelength [3]. Although there is not yet a full quantitative model for the detection and dark count mechanisms, the field has matured to the point where NbN devices are being tested in a variety of practical applications [4], [5]. Although NbN devices have been studied extensively, similar detectors made from other materials have not yet been studied in detail [6] - [8]. NbN is an attractive material for single photon detectors due to its short electron-phonon time. However, devices with an active area that is large enough to match typical optical systems have been shown to have a significantly longer reset time due to the nanowire's large kinetic inductance. Nb has a longer electron-phonon time but less kinetic inductance per unit length than NbN. Additionally, Nb is an easier material to process and has been studied extensively in the context of hot electron mixers and direct detectors [9], [10]. For these reasons, Nb is worth investigating as an alternative to NbN detectors.

In this paper, we investigate the performance of Nb nanowires used as photon counters. The devices tested consist of a nanostrip of Nb, 100 nm wide and 7.5 - 14 nm thick, on a sapphire substrate. The strip is patterned into a meander structure with area ranging from 4 to 100 μm² and pitch (center-to-center distance) of 200 nm (Figure 1) with fill factors of 50%. As in NbN devices, this forms a high aspect ratio wire with a typical width that is smaller than the magnetic penetration depth but larger than the coherence length, making it a quasi-one dimensional superconductor. During operation, the nanowire is biased with a dc current $I_b$ that is close to the measured critical current of the wire, $I_c$. The critical current is defined to be the value of dc current at which, when ramping up from zero current with no incident photons, the wire switches from the superconducting state into the resistive state. When the wire is biased near $I_c$ and a photon is absorbed, Cooper pairs are broken in a localized region, forming a hotspot. The hotspot disrupts the superconductivity across the wire, forming a resistive section with finite voltage, which can be detected with an rf amplifier. If the device is biased with a circuit that provides sufficiently strong negative electrothermal feedback, the hotspot will cool and superconductivity can be restored fully without active quenching or ramping down the dc bias current.

## II. METHODS AND PROCEDURE

The nanowires are fabricated using electron beam lithography and etching. Nb films are sputtered onto a 2" R-plane sapphire substrate at room temperature in an Ar plasma with a chamber base pressure of 5 x 10⁻⁹ torr. PMMA is spun on the wafer to an approximate thickness of 100 nm and patterned using an electron beam lithography system (Leica VB6 100 kV) and developed using a solution of IPA and water. The Nb film is patterned using a $CF_4$ reactive ion etch with the PMMA acting as an etch mask. We pattern the devices as well as the leads using a single lithography layer.

The data reported in this paper are from nanowires patterned from 14 and 8.5 nm thick films of Nb. Some of the 14 nm thick patterned nanowires were later thinned to approximately 7.5 nm using an argon ion beam, to test the effect of post process etching on device performance. Straight nanowires 10 μm in length as well as square meanders of 2 x

This work was supported in part by the National Science Foundation (EPDT) and by IBM Research. The principal author Anthony J. Annunziata is supported by a National Science Foundation Graduate Research Fellowship.

Anthony J. Annunziata, Daniel F. Santavicca, Joel D. Chudow, Luigi Frunzio, and Daniel E. Prober are in the Department of Applied Physics, Yale University, New Haven, CT 06520 (email: anthony.annunziata@yale.edu). Michael J. Rooks is with IBM Research at T. J. Watson Research Center in Yorktown Heights, NY 10598. Aviad Frydman is in the Department of Physics, Bar Ilan University, Ramat Gan, Israel.

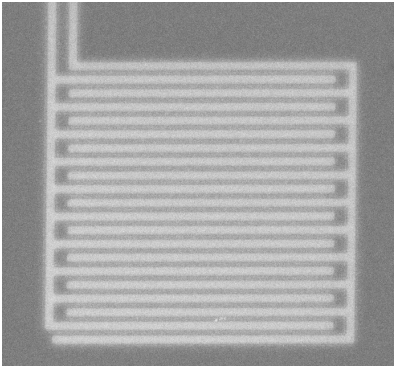

Fig. 1. SEM image of a 4x4 $\mu m^2$ Nb meander with coplanar waveguide in top left. The fill factor for this and all devices tested is 0.5; it appears less in this image due to substrate charging, which makes the gaps (bright) appear wider.

2, 4 x 4, and 10 x 10 $\mu m^2$ area were tested. Since we did not incorporate a capping layer, approximately 2 nm of the film thickness is assumed to be oxidized Nb and thus non-superconducting. This assumption is taken from previous work and from other reports on thin sputtered Nb [11]. Thickness measurements were made with an atomic force microscope (Digital Instruments Dimension 5000).

A simplified schematic of the readout circuit is shown in Figure 2 (a). The nanowire detector is wirebonded to a copper coplanar waveguide feeding into a coaxial input of a remote controlled 6-channel switch (Radiall R573423600 0-18 GHz), which enables testing 4 devices and 2 additional loads in any parallel combination during the same cooldown. DC and rf signals are coupled to the switch common port through a bias tee (Minicircuits ZFBT-6GW 0.1-6000 MHz, not shown) held at 4.2 K in the helium bath. RF amplification is accomplished using a cryogenic first stage amplifier (Amplitech APTC3-00050200-1500-P4) in the helium bath at 4.2K and a second stage amplifier at room temperature (Minicircuits ZFL1000LN 0.1 – 1000 MHz). There is a 6 dB attenuator at the input of the first stage amplifier (not shown) to eliminate standing waves due to impedance mismatches. Additional rf filters are also typically used (not shown) whose location and bandwidth depend on the measurement being performed. The dc line is filtered with a low pass copper powder filter (homemade, $f_{cutoff}$ ~ 1 MHz) held at 4.2 K in the helium bath. Optical excitation is through a multimode fiber (Ocean Optics ZFQ5426, 80 $\mu m$ core) where the cold end had been cleaved and left unconnectorized, and suspended approximately 3 mm above the four devices under test. The devices are located close to each other on the chip to ensure uniform light distribution (laser spot ≈ 1 mm diameter). The cryostat is a $^4$He system with a base temperature of ≈ 1.5 K. The cryostat is shielded with μ-metal and the sample is isolated from room light and electric fields by a copper inner vacuum can. DC biasing is from a low noise voltage source (Yokogawa 7651) in series with a large bias resistor (100 kΩ, 1 MΩ, or 10 MΩ). High speed readout is done using a 6 GHz, 20 gigasample/sec real time oscilloscope (Agilent 54855A).

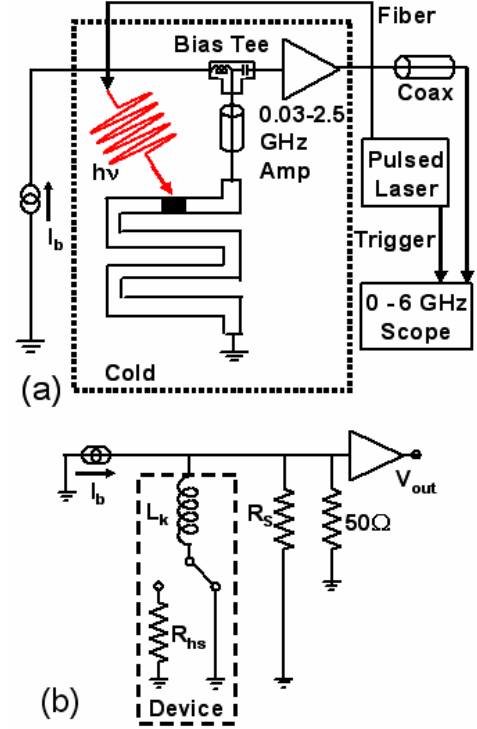

Fig. 2. (a) Simplified schematic of the measurement setup. (b) Equivalent circuit showing R = 0 before the photon is absorbed and after the hotspot has collapsed; $R_{hs}$ is the hot spot resistance after absorption and breaking of superconductivity; $R_s$ is a shunting resistor that can be added in parallel (at 4.2 K) to reduce the resistive load seen by the device.

### III. RESULTS

When the nanowire is biased with a current $I_b$ near but below $I_c$, the absorption of a single photon can produce an output voltage pulse. A typical pulse for a meander type detector with thickness $d \approx 14$ nm and line width $w = 100$ nm is shown in Figure 3 (inset). For all devices tested, the height of the pulse $V_p$ is $I_b \times R_{eff}$ where $R_{eff}$ is the total effective parallel resistance. The simplest case is when the extra shunt resistor $R_s = \infty$, for which $R_{eff}$ is just the input impedance of the microwave amplifier, $R_{in} = 50$ Ω. For finite $R_s$, we have $R_{eff} = 50 \times R_s / (50 + R_s)$. An equivalent circuit is seen in Figure 2 (b). The amplitude of the voltage pulse indicates that most of the bias current is transferred to the parallel circuit branch during detection, indicating the hotspot resistance is large compared to $R_{eff}$. Since this thick (14 nm) geometry yields the highest values for $I_c$, it typically has the highest signal-to-noise ratio; the detection efficiency is poor, however.

We used a highly attenuated pulsed laser source, with $\tau_{pulse}$ < 100 ps (FWHM) and repetition rate of 20 MHz (PicoQuant PDL-800b) to test the optical response. Since photons from a coherent source obey Poisson statistics, when the average photon number per pulse is much less than one, the count rate will scale linearly with laser power if the detector is single photon sensitive. Figure 3 shows count rate normalized by the laser repetition rate for 470 nm photons detected by a 10 x 10 $\mu m^2$ active area meander with line width of 100 nm and $d \approx 7.5$ nm made by thinning. Both thinned and directly sputtered devices were shown to be sensitive to single photons.

Detection efficiency is the fraction of photons incident on the total meander area (including gaps between strips) that produce a voltage pulse. The detection efficiency for the 7.5 nm devices is 6% +/- 3% for 470 nm wavelength photons and 1.2% +/- 0.6% for 690 nm photons when biased with $I_b$ near $I_c$. In comparison, the detection efficiency of directly sputtered 8.5 nm thin devices was only approximately 0.01% at 470 nm. The 14 nm thick devices have a detection efficiency of order $10^{-5}$%. Several sets of devices of each type were tested and detection efficiency was found to be similar as long as $I_C$ is similar (that is, there are no significant defects/constrictions; see [12]). The incident optical power is somewhat uncertain in our detection efficiency measurements due to interference effects within the multimode fiber.

The 7.5 and 8.5 nm films have similar superconducting properties (Table 1). The only significant difference is that 7.5 nm thinned devices have twice the resistivity. This has three effects. First, the inelastic electron-electron interaction time is shorter, which means more quasiparticles will be excited [13]. Second, the inelastic electron-phonon time is longer, which will limit the amount of heat lost to phonons, keeping more of the heat in the quasiparticle system [9]. Third, the diffusion length is shorter, resulting in more spatial confinement of the quasiparticles. The combined effect is to more strongly suppress the superconductivity in a smaller area, which should increase the probability that a resistive region will form across the wire. Thus, more resistive devices should in general have higher detection efficiency. More work is necessary to investigate the validity of this hypothesis, and to understand why the film resistivities differ.

Jitter was also measured by recording the standard deviation of the time delay between the laser trigger and the leading edge of the voltage pulse (measured at half height). We report an upper bound of 100 ps on the jitter of the thin Nb devices, limited by readout electronics and the finite width of the laser pulse in time.

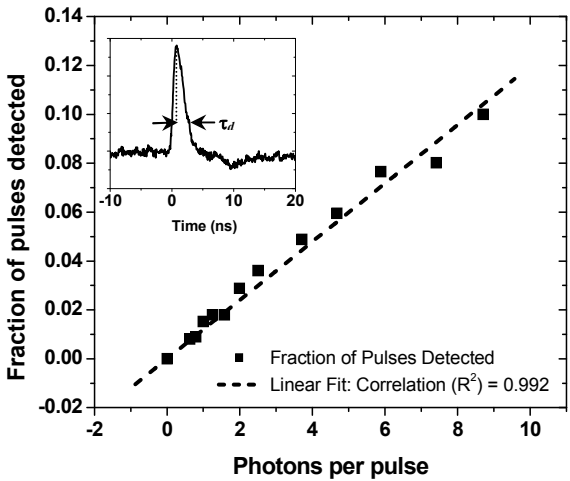

Fig. 3. A linear relationship in detection rate (count rate divided by laser repetition rate) versus average number of photons incident on the detector demonstrates single photon sensitivity. Data taken for 10 x 10 μm² Nb meander type detector with $w$ = 100 nm, $d$ = 7.5 nm (thinned), $T_c$ = 4.5 K, $I_c$ = 8.2 μA, and $R_{eff}$ = 25 Ω (see text). Inset: typical voltage pulse for single photon detection; the decay time is the $1/e$ time obtained by fitting a decaying exponential function to the falling edge of the pulse; $\tau_{reset} \approx 3 \times \tau_d$.

As in NbN, the detection efficiency of all Nb nanowire detectors depends strongly on how close they are biased to $I_c$. For high detection efficiency, the ideal bias range is $I_b/I_c > 0.9$. The ability to bias close to $I_c$, however, is impaired by thermal instability. For the device to recover after a detection event (that is, for the hotspot to collapse back to the fully superconducting state), the circuit that is used to read out the device must also supply negative electrothermal feedback. This is well understood in other types of superconducting detectors, e.g., superconducting transition edge sensors (TESs), which are dc voltage biased in a finite voltage state with a hotspot that varies in size (resistance) in response to absorbed radiation [14]. In a TES, the hotspot is perturbed only mildly by incident radiation. A nanowire detector is perturbed strongly by an absorbed photon. In both cases, the return to equilibrium is dependent on the readout circuit. Our data below show that Nb devices are much more susceptible to thermal runaway than are NbN devices when using a 50 Ω amplifier.

To investigate thermal runaway in detail, we have added various values of shunt resistance to the 50 Ω readout circuit to lower the total effective parallel impedance seen by the device at frequencies corresponding to the reset dynamics (typically 30 MHz – 3 GHz). The amplifier rf input impedance is 50 Ω (real). The magnitude of the impedance of the dc bias branch of the circuit is large enough to be neglected. We add a resistor, $R_s$, in parallel with the coaxial input but located close to the device to minimize the effect of standing waves (see Figure 2 (b)). We first test with $R_s = \infty$, for which $R_{eff} \approx 50\ \Omega$. For this case we find that at currents $I_b \geq 0.6I_c$, the Nb device will not reset to the fully superconducting state after detecting a photon. Rather, the hotspot grows until the nanowire has fully transitioned to the normal state and remains there. We say that such a device has latched, and we define the current at which the device latches as $I_{latch}$. Due to the self-heating hysteresis in the dc IV curve, the device will remain latched until the dc bias current is lowered below the point at which the dc heating is overcome by the cooling of the nanowire; this occurs at $I_{return,\ dc}$. In general, $I_{latch} > I_{return,\ dc}$.

The data show that with a lower value of $R_{eff}$, the Nb devices can be biased closer to $I_c$. A summary of the stable biasing range (where the device does not latch) is seen in Figure 4 for a 10 x 10 μm², $d \approx 7.5$ nm thinned device. The effect of lower $R_{eff}$ is to maintain a more negative feedback condition (closer to an ac voltage bias where the dissipated power, $P_{dissipated} = V^2/R_{hotspot}$ with $dP = -V^2/R_{hs}^2 dR$) for a large fraction of the hotspot lifetime. The smaller value of $R_{eff}$ also slows down the return of the current from the amplifier branch back into the device. The value of $R_{eff}$ at which the nanowire can be biased with $I_b > 0.9I_c$ depends on the device sheet resistance and total length. Thus, the numerical scale of the x-axis of Figure 4 is not universal for all detector geometries and parameters. In general, $R_{eff}$ must be smaller for shorter devices with lower inductance and for devices with lower sheet resistance. In comparison, for NbN devices, $R_{sheet}$ is a factor of 10-20 greater than for Nb so $R_{eff} = 50\ \Omega$ is small enough for stable operation of all but the shortest NbN detectors. The shortest detectors have the lowest inductance and fastest reset time [15].

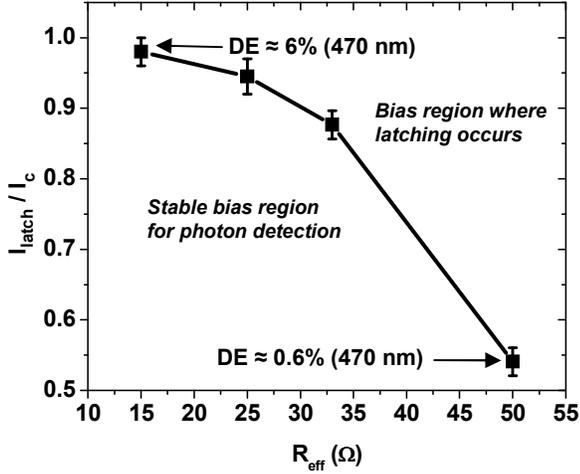

Fig. 4. Latching current versus effective parallel resistance ($R_{eff} = R_{in} \parallel R_s$ amplifier in parallel with $R_s$). Error bars are due to variation in measured latching current over multiple measurements and not uncertainty in individual measurements. Data taken for the same device as in Figure 3.

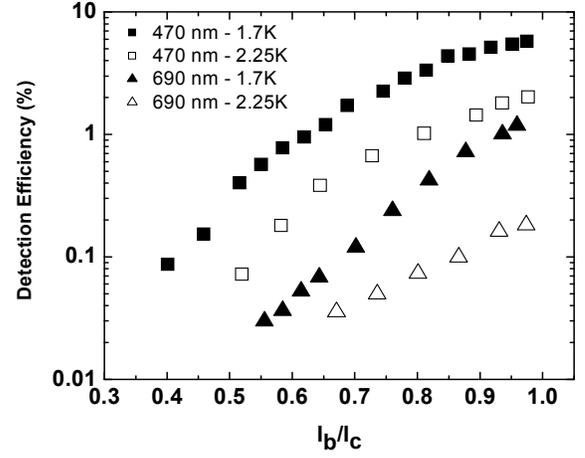

Fig. 5. Detection efficiency (DE) dependence on bias current at 1.7 K and 2.25 K for 470 nm and 690 nm photon wavelengths. Data taken for the same device as in Figure 3 with $R_{eff} = 25\ \Omega$; dark counts have been subtracted.

With the lowest value of $R_{eff}$, latching is avoided and we can measure how the detection efficiency depends on bias current nearly all the way up to $I_b \approx I_c$. Figure 5 shows this dependence plotted versus $I_b$ for the same device as measured in Figure 4. Here, $R_s = 50\ \Omega$ so that $R_{eff} = 25\ \Omega$. As in NbN, the detection efficiency scales exponentially with $I_b$ for small $I_b$, approximately saturating at high bias when most of the absorbed photons are detected (the fraction of the incident flux that is absorbed is small) [3]. Also as for NbN, the detection efficiency improves at lower temperature when holding the fractional bias current fixed. Thus, empirically the detection mechanism in Nb is qualitatively similar to NbN. Further work is needed to understand this dependence.

We find that the decay time, $\tau_d$, of Nb detectors is given by $L_{total}/R_{eff}$ for long (>50 μm) nanowires. For all geometries tested, our measurements confirm that the total inductance, $L_{total}$, is dominated by the kinetic inductance of the supercurrent, which is proportional to the length of the nanowire. Simple theoretical considerations from the one-dimensional Ginzberg-Landau theory give $L_K = \mu_0 \lambda^2 \times \ell/A_{cs}$ where $\lambda$ is the magnetic penetration depth, $\ell$ is the nanowire length, and $A_{cs}$ is the cross sectional area of the nanowire [16]. Using typical values for the devices with the best detection efficiency (7.5 nm ion beam thinned), this predicts that Nb devices should have a factor of 3 less kinetic inductance than NbN for the same geometry nanowire. Direct measurements of $L_K$ were made with a microwave network analyzer by incorporating the devices into a resonant circuit which confirm these predictions. However, low values of $R_{eff}$ are necessary to avoid latching at high bias ($I_b > 0.9I_c$). For thinned devices, $R_{eff}$ must be approximately 30 Ω or less. Thus, the pulse decay time for long devices, $\tau_d = L_K/R_{eff}$, is only a factor of approximately 1.5 smaller than NbN. The shortest decay time for a 10 x 10 μm² thinned Nb device is then 6 ns since $L_K$ for these most resistive devices is 0.45 nH/μm. For shorter Nb devices, the decay time is faster (since total $L_K$ is lower) but even the shortest devices had $\tau_d > 1.5$ ns when using an $R_{eff}$ that enabled biasing at $I_b > 0.9I_c$. In general, the decay time is well described by the equation $\tau_d = \tau_0 + L_K/R_{eff}$, with $\tau_0 \approx 1$ ns (close to the measured intrinsic thermal cooling time in Nb [10]) and the optimal value of $R_{eff} \approx 20\ \Omega$ for the shortest devices. The maximum count rate of these devices is approximately $(3 \times \tau_d)^{-1}$ since the device current must return very near to the dark value before another photon is detected.

TABLE I: DEVICE PARAMETER COMPARISON

| Parameter (typical values) | Thinned | Thin direct deposition | Thick direct deposition |
|---|---|---|---|
| $d$ | 7.5 nm | 8.5 nm | 14 nm |
| $w$ | 100 nm | 100 nm | 100 nm |
| $R_{sheet}$ (10 K) | 105 Ω/square | 45 Ω/square | 18 Ω/square |
| $T_c$ | 4.5 K | 4.3 K | 6.5 K |
| $I_c$ | 8.2 μA | 7.4 μA | 20 μA |
| Det.Eff. at 470 nm | 6 % | $\approx 10^{-2}$ % | $\sim 10^{-5}$ % |

Table I: Thinned refers to devices that were ion milled to 7.5 nm from 14 nm. Thin and thick direct deposition refer to devices that were patterned directly in 8.5 nm or 14 nm films. The typical value of a parameter is reported, obtained from measuring several devices fabricated in the same way.

## IV. CONCLUSION

We have fabricated and tested Nb nanowire single photon detectors and shown that the best devices have detection efficiency of approximately 6% at 470 nm. We have demonstrated that Nb devices are much more susceptible to thermal latching than NbN devices with a 50 Ω amplifier. We have provided a solution to this problem that entails reducing the effective input resistance of the readout network to be less than 50 ohms. This increases the strength of the negative electrothermal feedback, stabilizing the device at high bias. In light of this, the reset times of Nb devices are approximately a factor of 1.5 shorter than NbN for similar geometry detectors.